\documentclass[pdflatex,sn-mathphys-num]{sn-jnl}% Math and Physical Sciences Numbered Reference Style
\usepackage{graphicx}%
\usepackage{multirow}%
\usepackage{amsmath,amssymb,amsfonts}%
\usepackage{amsthm}%
\usepackage{mathrsfs}%
\usepackage[title]{appendix}%
\usepackage{xcolor}%
\usepackage{textcomp}%
\usepackage{manyfoot}%
\usepackage{booktabs}%
\usepackage{algorithm}%
\usepackage{algorithmicx}%
\usepackage{algpseudocode}%
\usepackage{listings}%
\usepackage{lineno}  
\usepackage{rotating} %  sidewaysfigure / sidewaystable
\usepackage[dvipsnames]{xcolor} 
\usepackage{soul}               
\sethlcolor{Goldenrod}

\usepackage{subcaption}

\theoremstyle{thmstyleone}%
\theoremstyle{thmstyletwo}%

\theoremstyle{thmstylethree}%
\usepackage{comment}
\newcommand{\coSeniorMark}{\textsuperscript{\ddag}}
\begin{document}
%\pagenumbering{gobble}
\pagenumbering{arabic}
%\linenumbers  
\title[Article Title]{Interpretable and backpropagation-free Green Learning for efficient multi-task echocardiographic segmentation and classification}

\author[1,2]{\fnm{Jyun-Ping} \sur{Kao}}\email{jjpkao@gmail.com}
\equalcont{These authors contributed equally to this work.}
\author[3]{\fnm{Jiaxin} \sur{Yang}}\email{yangjiax@usc.edu}
\equalcont{These authors contributed equally to this work.}

\author[3]{\fnm{C.-C. Jay} \sur{Kuo}\coSeniorMark}\email{jckuo@usc.edu}
\author*[1]{\fnm{Jonghye} \sur{Woo}\coSeniorMark}\email{jwoo@mgh.harvard.edu}

\begingroup
\renewcommand\thefootnote{\ddag}
\footnotetext{These authors are the senior authors.}
\endgroup

\affil[1]{\orgdiv{Department of Radiology}, \orgname{Harvard Medical School and Massachusetts General Hospital}, \orgaddress{\city{Boston}, \state{MA}, \country{USA}}}

\affil[2]{\orgdiv{Graduate Institute of
Biomedical Electronics and Bioinformatics}, \orgname{National Taiwan University}, \orgaddress{\city{Taipei}, \country{Taiwan}}}

\affil[3]{\orgdiv{Ming Hsieh Department of Electrical and Computer Engineering}, \orgname{University of Southern California}, \orgaddress{\city{Los Angeles}, \state{CA}, \country{USA}}}

\abstract{
Echocardiography is a cornerstone for managing heart failure (HF), with Left Ventricular Ejection Fraction (LVEF) being a critical metric for guiding therapy. However, manual LVEF assessment suffers from high inter-observer variability, while existing Deep Learning (DL) models are often computationally intensive and data-hungry ``black boxes" that impede clinical trust and adoption. Here, we propose a backpropagation-free multi-task Green Learning (MTGL) framework that performs simultaneous Left Ventricle (LV) segmentation and LVEF classification. Our framework integrates an unsupervised VoxelHop encoder for hierarchical spatio-temporal feature extraction with a multi-level regression decoder and an XG-Boost classifier. On the EchoNet-Dynamic dataset, our MTGL model achieves state-of-the-art classification and segmentation performance, attaining a classification accuracy of 94.3\% and a Dice Similarity Coefficient (DSC) of 0.912, significantly outperforming several advanced 3D DL models. Crucially, our model achieves this with over an order of magnitude fewer parameters, demonstrating exceptional computational efficiency. This work demonstrates that the GL paradigm can deliver highly accurate, efficient, and interpretable solutions for complex medical image analysis, paving the way for more sustainable and trustworthy artificial intelligence in clinical practice.
}


\maketitle

%\newpage
%==== Done======↓
\section{Introduction}\label{sec1}
Cardiovascular echocardiography is a non-invasive diagnostic imaging modality that uses high-frequency ultrasound waves to produce real-time, dynamic images of cardiac structures, myocardial function, and intracardiac hemodynamics~\cite{borlaug2020evaluation, ziaeian2016epidemiology}. It has become a foundational pillar in modern clinical cardiology \cite{borlaug2020evaluation, ziaeian2016epidemiology}.

The clinical utility of echocardiography is exceptionally broad, spanning the entire spectrum of cardiovascular disease \cite{omar2016advances, marwick2015role}. In the management of heart failure (HF), it is an indispensable tool for establishing the diagnosis, identifying the underlying etiology, and providing comprehensive assessments of systolic and diastolic function, chamber dimensions, and valvular integrity \cite{marwick2015role}. Clinical guidelines recommend echocardiography for all patients with suspected HF, and studies have shown that its use is associated with improved adherence to evidence-based therapies and superior survival outcomes \cite{heidenreich20222022, maddox20242024}.

A cornerstone of this assessment is the quantification of left ventricular ejection fraction (LVEF), a fundamental measure of global systolic function and a powerful predictor of outcomes \cite{redfield2023heart, cole2015defining}. Clinical guidelines rely heavily on precise LVEF thresholds to classify HF into distinct phenotypes: HF with Reduced Ejection Fraction (HFrEF; $\text{LVEF} < 40\%$), HF with Mid-range Ejection Fraction (HFmrEF; $40\% \le \text{LVEF} \le 50\%$), and HF with Preserved Ejection Fraction (HFpEF; $\text{LVEF} > 50\%$) \cite{redfield2023heart, cole2015defining}. These classifications directly inform crucial therapeutic decisions, such as eligibility for implantable cardioverter-defibrillators or specific pharmacotherapies \cite{glikson2001implantable, turgeon2021pharmacotherapy}.

Yet, the clinical utility of LVEF is undermined by limitations in either its manual tracing or in the verification of semi-automated measurements~\cite{klaeboe2019echocardiographic, knackstedt2015fully}.The conventional method, which requires manual tracing of the endocardial border, is not only laborious but also inherently subjective, resulting in substantial inter- and intra-observer variability~\cite{lui-intra, kouris2021left}. Such imprecision is clinically perilous: a variability of this magnitude can misclassify patients and lead to incorrect management, fundamentally challenging the reliability of evidence-based care when measurement error rivals the width of the diagnostic categories themselves \cite{christersson2024usefulness, kouris2021left}.

Over the past decade, deep learning (DL) has revolutionized medical image analysis. Convolutional neural networks (CNNs), such as the U-Net architecture~\cite{unet}, have become the standard for image segmentation, and increasingly complex models continue to push performance boundaries \cite{azad2024medical}. In echocardiography, the video-based DL model EchoNet-Dynamic \cite{ouyang2020video}, trained on over 10,000 annotated images, was shown to segment the LV with human-expert accuracy and estimate EF within a few percentage points of ground truth for HF classification. These results demonstrate the potential of modern CNN-based algorithms to augment clinical echo interpretation. However, significant computational and practical challenges persist because state-of-the-art CNNs typically demand large labeled datasets and extensive training via backpropagation \cite{alzubaidi2021review}. In medical imaging domains such as echocardiography, data acquisition and expert annotation are labor-intensive, a factor that often limits dataset size~\cite{yoon2021artificial}. Moreover, DL models with millions of parameters can be inefficient to deploy on edge devices and lack interpretability, rendering their decision-making processes an effective `black box' that clinicians may not fully trust \cite{kuo2016understanding, GL}. These issues motivate the search for more efficient, interpretable, and resource-light solutions for segmentation and classification using echocardiography.

Recently, Green Learning (GL)~\cite{GL} has emerged as a promising alternative paradigm to address the shortcomings of conventional DL models. In contrast to large backpropagation-trained networks, GL methods employ feed-forward, modular architectures that do not require backpropagation at all. They are characterized by drastically lower model complexity and energy consumption, using analytical or data-driven subspace learning in place of end-to-end gradient descent. As a result, GL models tend to have small parameter counts, low computational footprints, and logically transparent operations. These properties align well with the needs of clinical AI systems, which often require running on limited hardware and providing human-interpretable results \cite{chen2022explainable}. By avoiding iterative weight tuning, GL techniques can also mitigate the risk of overfitting on small datasets. Early demonstrations of the GL paradigm have shown that feed-forward designs can achieve performance on par with deep CNNs in vision tasks while being far more lightweight. For example, PixelHop \cite{pixelhop}, PointHop \cite{pointhop}, and VoxelHop \cite{voxelhop} are successive subspace learning methods that extract multi-stage features using statistical transforms, such as Principal Component Analysis (PCA), achieving competitive results without the need for backpropagation. This “green” machine learning approach is especially attractive for medical applications where interpretability and efficiency are of paramount importance. Building on this, the Green U-Shaped Learning (GUSL) \cite{GUSL} model was introduced as one of the first GL-based frameworks for medical image segmentation, demonstrating state-of-the-art performance on prostate MRI segmentation with a fraction of the parameters of its DL counterparts. It employs successive linear feature extraction and per-pixel regression modules that mirror the “U-shape” of a traditional U-Net~\cite{unet}. Notably, this feed-forward design provides inherent interpretability: features are computed via linear transforms whose contributions can be traced, and the model can explicitly focus on difficult regions by modeling the residuals.

While DL has delivered strong performance in echocardiography, practical deployment presents complexities. Modern compression strategies, including pruning, quantization, distillation, and lightweight backbone design, can substantially reduce latency and memory, enabling point-of-care inference on portable devices~\cite{narang2021utility}. Nevertheless, many high-performing pipelines still depend on backpropagation-intensive training, large labeled datasets, and limited transparency in learned representations, which complicates governance, debugging, and clinical trust, particularly when models must be updated across new scanners, protocols, or patient populations~\cite{ouyang2020video, amann2020explainability, GL}.

We hypothesize that a backpropagation-free, multi-task Green Learning model using an unsupervised VoxelHop encoder and regression/gradient-boosted decoders can achieve segmentation and LVEF-classification performance comparable to or better than state-of-the-art 3D DL models, while requiring substantially fewer parameters and improving interpretability.

Building upon these insights, we develop a novel end-to-end interpretable multi-task framework for echocardiographic analysis that addresses both segmentation and cardiac function classification. For segmentation, we adopt GUSL’s \cite{GUSL} strategy of using a VoxelHop-based encoder paired with a coarse-to-fine residual regression decoder. For classification, we integrate a feature-selecting XGBoost classifier \cite{xgboost} on the learned representations. Notably, no component of our model requires backpropagation: the feature encoder is constructed via Principal component analysis (PCA), the segmentation regressors are solved in closed form (least squares), and the XGBoost classifier uses gradient-boosted decision trees. The entire pipeline is trained in a layer-wise, modular fashion and has orders of magnitude fewer trainable parameters than typical DL models.

%==== Done======↑

\section{Materials and Methods}\label{sec3}

In this section, we outline the methodology for LV segmentation and LVEF classification on echocardiograms using our multi-task Green Learning (MTGL) model. The approach consists of a unified preprocessing strategy (Section ~\ref{preprocess}), followed by two task-specific pipelines, including a segmentation pipeline (Section ~\ref{seg-encoder}) based on the GUSL~\cite{GUSL} architecture for volumetric segmentation, and a classification pipeline (Section ~\ref{class-branch}) that leverages features extracted from the GUSL encoder for LVEF classification.

\subsection{Data and Preprocessing}\label{preprocess}

\begin{figure}[h]
\centering
\includegraphics[width=0.9\textwidth]{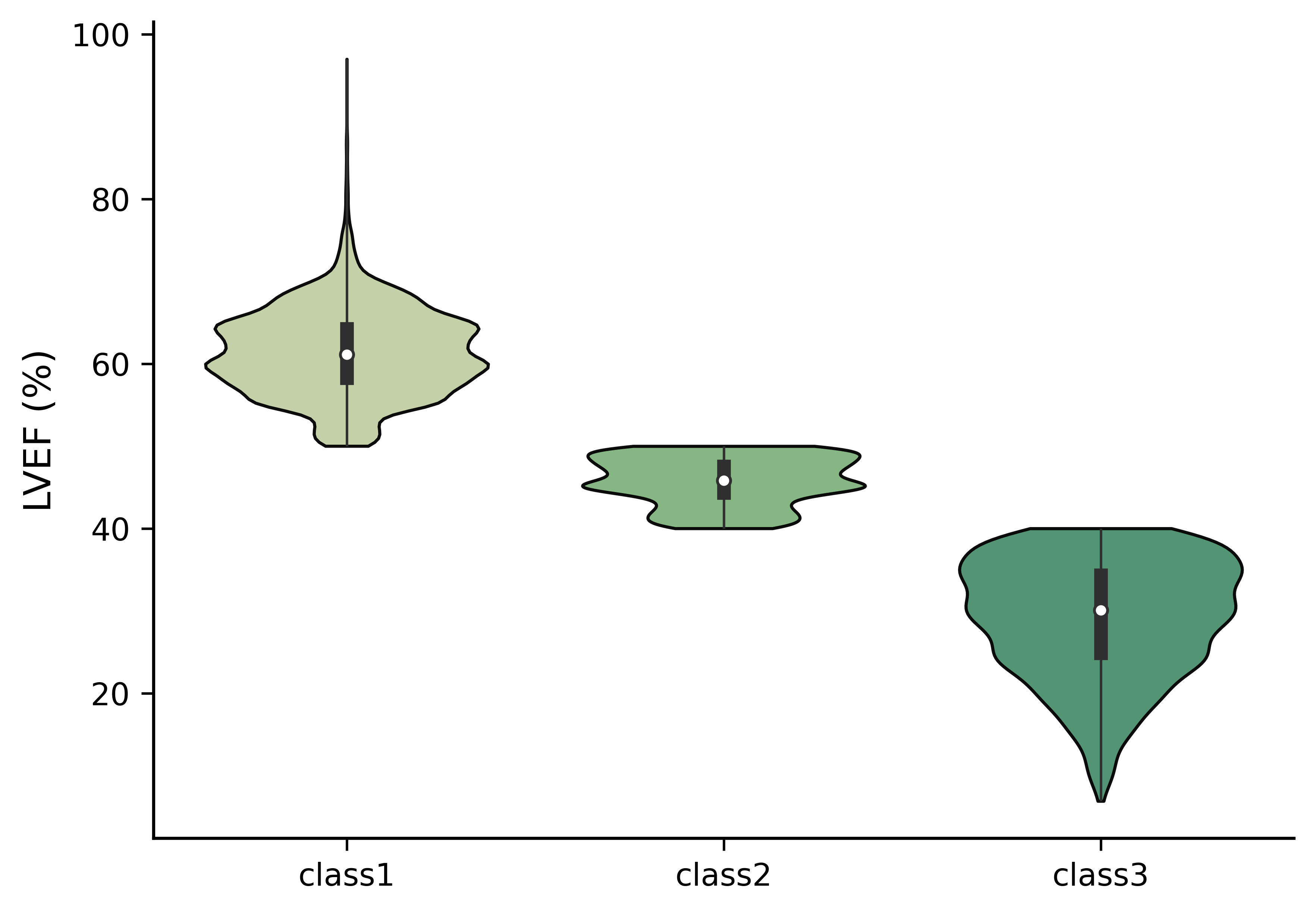}
\caption{Data Distribution of classification task. Class 1 denotes as LVEF $>$ 50\%, class 2 denotes as LVEF between 40–50\% and class 3 denotes as LVEF between $<$ 40\%.}\label{EF_by_class}
\end{figure}

We used the open-source echocardiogram dataset provided by EchoNet-Dynamic \cite{ouyang2020video}. This dataset comprises 10,030 echocardiography videos obtained at Stanford Health Care from 10,030 distinct individuals who underwent echocardiography between 2016 and 2018. The videos were randomly partitioned into training (n = 7,465), validation (n = 1,277), and test (n = 1,288) sets. Each frame and its corresponding mask are resized to $128\times128$ pixels. For each frame, we stack the resized End-Diastolic Volume (EDV) and End-Systolic Volume (ESV) images from the same video series of size $112\times112\times2$ as a single dataset. Thus, each pixel position now contains a 2-dimensional feature: [EDV, ESV] (one cardiac cycle). Ground truth segmentations were derived from the EchoNet-Dynamic dataset, which provides expert coordinate tracings of the LV at end-diastole and end-systole; these were converted into binary masks to train and evaluate the segmentation decoder. We selected a sequence of 12 consecutive frames from each video and concatenate their two-channel frames along the temporal axis to serve as the input to the MTGL model, and we followed the train/validation/test split defined in the EchoNet-Dynamic dataset \cite{ouyang2020video} for all experiments. For the classification task, the patient’s LVEF is discretized into three classes based on clinical cut-offs derived from the expert-verified manual tracings provided in the dataset: (i) EF $>$ 50\% (normal systolic function), (ii) EF in 40–50\% (mildly reduced), and (iii) EF $<$ 40\% (significantly reduced) \cite{redfield2023heart, cole2015defining} and the data distribution are shown in Figure~\ref{EF_by_class}.

%Our segmentation method enhanced the GUSL model originally proposed for 3D MRI prostate segmentation to the task of LV segmentation in 2D echocardiography videos \cite{GUSL}. GUSL is a feed-forward, all-regression segmentation framework that forgoes backpropagation-based training in favor of an interpretable, two-stage pipeline of unsupervised feature learning and multi-level regression. This model is “U-shaped” in the sense that it performs multi-scale analysis and subsequent coarse-to-fine mask prediction refinement, analogous to a U-Net, but uses statistical learning, including PCA and regression at each stage, instead of deep neural network layers. Our model could break down into an unsupervised VoxelHop encoder (Section ~\ref{vox-encoder}), a segmentation decoder (Section ~\ref{seg-encoder}), and a classification decoder (Section ~\ref{class-branch}). The encoder is pre-trained once via unsupervised learning and subsequently kept fixed (no gradient updates) during downstream training. The segmentation and classification decoders are then trained independently on their respective datasets. During the inference stage, the pipeline operates end-to-end: an input is processed by the same encoder, and its representation is routed to the corresponding decoder to yield a segmentation mask and a class label simultaneously.

\begin{figure}[h]
\centering
\includegraphics[width=0.9\textwidth]{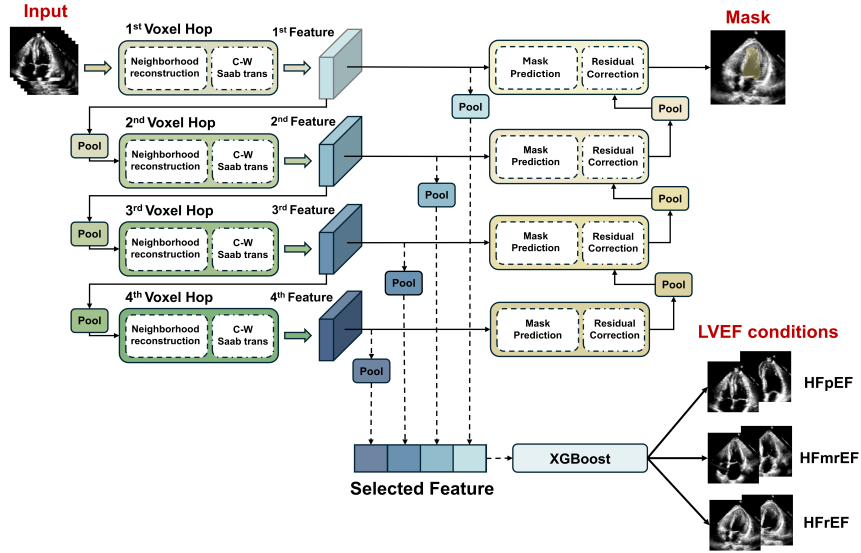}
\caption{Multi-task Green Learning (MTGL) Model Architecture.}\label{model}
\end{figure}

Our segmentation method enhanced the GUSL model~\cite{GUSL} originally proposed for 3D MRI prostate segmentation to the task of LV segmentation in 2D echocardiography videos, which is a feed-forward, all-regression segmentation framework that forgoes backpropagation-based training in favor of an interpretable, two-stage pipeline of unsupervised feature learning and multi-level regression. Our model (Figure~\ref{model}) consists of an unsupervised VoxelHop encoder (Section ~\ref{vox-encoder}), a supervised segmentation decoder (Section ~\ref{seg-encoder}), and a supervised classification decoder (Section ~\ref{class-branch}). The encoder is pre-trained once via unsupervised learning and subsequently kept fixed during downstream training. The segmentation and classification decoders are then trained independently on their respective datasets. During the inference stage, the pipeline operates end-to-end: an input is processed by the same encoder, and its representation is routed to the corresponding decoder to yield a segmentation mask and a class label simultaneously. For benchmarking, we implemented 3D V-Net, 3D U-Net, 3D UNETR, and 3D nnU-Net baselines using the same input volumes, preprocessing, data splits, and evaluation protocol as our method to enable a fair comparison.

\subsubsection{Unsupervised VoxelHop Encoder}\label{vox-encoder}
%In the unsupervised voxelHop encoder, we extract multi-scale features from the input volume using four cascaded VoxelHop units \cite{voxelhop}. VoxelHop is a 3D extension of the PixelHop Saab transform method \cite{pixelhop}, which learns feature filters through successive PCA rather than gradient descent, diminishing the need for back-propagation. The input volume is fed into a hierarchy of four VoxelHop layers, each of which expands the receptive field and produces a richer representation of each voxel in an unsupervised manner. 

In the unsupervised VoxelHop encoder, we extract multi-scale features from the input volume using four cascaded VoxelHop units~\cite{voxelhop}, which learns feature filters through successive PCA~\cite{PCA} rather than gradient descent, diminishing the need for back-propagation. The input volume is fed into a hierarchy of four VoxelHop layers, each of which expands the receptive field and produces a richer representation of each voxel in an unsupervised manner.

In each VoxelHop layer, the following operations occur: 
\begin{itemize}
\item Neighborhood construction: a 3D spatial cuboid around each voxel is extracted, aggregating local spatial-temporal context; 
\item Channel-wiseSaab Subspace approximation via adjusted bias (Saab) transform: a channel-wise PCA is applied to this neighborhood to produce decorrelated spectral features. 
Each VoxelHop encoder adopts the Saab transform at each hop to obtain an interpretable, linear decomposition of a local 3D neighborhood into one direct current (DC) component and multiple alternating current (AC) components.
Let $\mathbf{x}\in\mathbb{R}^{d}$ denote the vectorized $s\times s\times k$ cuboid centered at a voxel. Saab uses a fixed DC anchor
$\mathbf{a}_0=\frac{1}{\sqrt{d}}\mathbf{1}$ and an adjusted bias $b$ so that the DC response
$y_0=\mathbf{a}_0^\top\mathbf{x}+b$ captures the local mean intensity. The AC subspace is the orthogonal complement of $\mathrm{span}(\mathbf{a}_0)$, on which PCA is applied to yield orthonormal AC filters $\{\mathbf{a}_j\}_{j=1}^{d-1}$ with eigenvalues $\{\lambda_j\}_{j=1}^{d-1}$ ordered as $\lambda_1\ge \lambda_2\ge\cdots$. The AC responses $\{y_j=\mathbf{a}_j^\top\mathbf{x}\}$ encode mean-removed, decorrelated directional variations of the neighborhood (edges, textures, motions), while $y_0$ carries low-frequency content. This DC/AC split makes the encoder a ``white-box" in which each channel has a clear algebraic and signal-processing meaning, and the contribution of any subset of AC filters can be quantified by its eigen-energy.

\end{itemize}

%In each VoxelHop layer, the following operations occur: 
%\begin{itemize}
%\item Neighborhood construction: a 3D spatial cuboid around each voxel is extracted, aggregating local spatial-temporal context.
%\item Channel-wiseSaab Subspace approximation via adjusted bias (Saab) transform: a channel-wise PCA \cite{PCA} is applied to this neighborhood to produce decorrelated spectral features. 
%Each VoxelHop encoder adopts the Saab transform at each hop to obtain an interpretable, linear decomposition of a local 3D neighborhood into one direct current (DC) component and multiple alternating current (AC) components. This DC/AC split makes the encoder a ``white-box" in which each channel has a clear algebraic and signal-processing meaning, and the contribution of any subset of AC filters can be quantified by its eigen-energy.

%\end{itemize}

After each VoxelHop, a 3D Max-Pooling layer is used to reduce spatial resolution by a factor of 2, progressively coarsening the feature map dimensions.
As a result, we obtain four feature outputs from \textit{Hop~1} through \textit{Hop~4}. Importantly, the VoxelHop training does not require labels; it is entirely unsupervised and data-driven, yielding an interpretable linear feature representation for subsequent regression. 
The kept set of AC filters provides an interpretable solution for the entire encoder. First, the monotone curve, rather than opaque regularizers or training heuristics, serves as a transparent certificate that each hop meets the target energy level and ensures keeping only useful features in the vector to significantly lower the training parameters (Fig.~\ref{fig-energy}). Because energy spectra differ across hops, selecting ACs per hop avoids a one-size-fits-all channel count. Moreover, each Hop's AC filters could be visualized after training, enabling us to understand ``What" and ``How" the model learned, helping us monitor the training process and clearly provide directions for improving the model.
Hence, the encoder exposes a clear and interpretable linear feature space to the subsequent encoder branch, aligning with the white-box philosophy.

%==== Done======↑

\subsubsection{Segmentation decoder}\label{seg-encoder}
%The segmentation decoder stage of multi-task GUSL produces the segment mask via a coarse-to-fine residual correction strategy, following GUSL's pipeline \cite{GUSL}. However, it removes the 2-stage ROI selection segmentation due to the relatively low resolution of the input data, thereby lowering the training complexity. Rather than directly predicting a high-resolution mask in one step, multi-task GUSL trains a sequence of regression models at different scales, where each model corrects the errors of the previous (coarser) level. All regression models in our pipeline are implemented using XGBoost \cite{xgboost}, chosen for its efficiency and ability to handle large numbers of features. We denote the four resolution levels from coarsest (level 4) to finest (level 1):

The segmentation decoder stage of MTGL produces the segment mask via a coarse-to-fine residual correction strategy~\cite{GUSL}. However, we remove the 2-stage ROI selection segmentation due to the relatively low resolution of the input data, thereby lowering the training complexity. Rather than directly predicting a high-resolution mask in one step, MTGL trains a sequence of regression models at different scales, where each model corrects the errors of the coarser level. All regression models in our pipeline are implemented using XGBoost \cite{xgboost}, chosen for its efficiency and ability to handle large numbers of features.
We denote the four resolution levels from coarsest (level 4) to finest (level 1):

\begin{enumerate}[1.]
\item Level 4 (Coarsest) – Initial Mask Prediction: \\
At the lowest resolution (level 4), we train an XGBoost regressor to predict a coarse mask for the LV. The regressor takes as input the VoxelHop features for each voxel at this scale and outputs a continuous-valued score between 0 and 1, representing the probability that the voxel belongs to the LV region. To obtain training targets for this regressor, we downsample the ground-truth binary masks to the level 4 resolution using patch averaging. This means that each target value at level 4 is the fraction of the corresponding $8\times8$ block of the original image that is occupied by the LV (a real number in the range [0, 1]). Using these fractional labels preserves sub-pixel boundary information during training, allowing the XGBoost model to treat segmentation as a voxel-wise regression problem. The initial XGBoost is trained on all voxels within the region of interest of the image. In practice, since much of the echo frame consists of background, we first crop the input volume to a tight region around the heart to balance the foreground-to-background sample ratio. This yields a coarse segmentation output at 14$\times$14 resolution.

At the lowest resolution (level 4), we train an XGBoost \cite{xgboost} regressor to predict a coarse mask for the LV. The regressor takes as input the VoxelHop features for each voxel at this scale and outputs a continuous-valued score between 0 and 1, representing the probability that the voxel belongs to the LV region. To obtain training targets for this regressor, we downsample the ground-truth binary masks to the level 4 resolution using patch averaging. This means that each target value at level 4 is the fraction of the corresponding $8\times8$ block of the original image that is occupied by the LV. Using these fractional labels preserves sub-pixel boundary information during training, allowing the XGBoost model to treat segmentation as a voxel-wise regression problem. In practice, since much of the echo frame consists of background, we first crop the input volume to a tight region around the heart to balance the foreground-to-background sample ratio. This yields a coarse segmentation output at 14$\times$14 resolution.

\item Residual Error Computation and ROI Attention: \\
After obtaining the initial prediction at level 4, we compute the residual error map, defined as the difference between the predicted mask and the downsampled ground truth mask at that scale. We observe that this residual is largely concentrated near the boundaries of the ventricle, since interior voxels (entirely inside LV) and exterior voxels (entirely outside LV) are often already predicted accurately, whereas voxels overlapping the true boundary tend to have fractional errors. To focus the model on these error-prone boundary regions, we employ an ROI-based sampling strategy. Specifically, we partition the voxels of the coarse prediction into three groups: background, LV interior, and boundary (transition) region, based on the ground truth. The residual map has near-zero values for most background/interior voxels, but higher values along the boundary. We therefore define an ROI that encompasses primarily the boundary voxels and down-sample the training samples to include mainly those voxels. A second XGBoost model (a residual regressor) is then trained on this ROI-focused subset to predict the residual error values at level 4. In essence, this model “attends” to the LV edges and learns to correct the under- or over-segmentation in those regions. By applying this residual correction, we obtain an improved coarse mask at level 4.

\item Levels 3, 2, and 1 – Up-sampling and Residual Correction: \\
The refined coarse prediction is successively propagated to higher resolutions. We up-sample the level 4 prediction after residual correction back to the resolution of level 3. At level 3, another XGBoost regression model is trained to predict the residual between the up-sampled mask and the ground truth mask, downsampled to level 3 resolution. We again concentrate on ROI voxels (predominantly boundary areas) for this residual training. The predicted residual at level 3 is added to the upsampled mask to achieve a more accurate segmentation at that scale. This process is repeated for level 2 and then for level 1 (full $112\times112$ resolution): at each stage, the current prediction is interpolated to the next finer scale, and an XGBoost model corrects the discrepancies by learning the residual relative to the ground truth at that scale. By level 1, the model outputs a high-resolution probability map of the LV.

\end{enumerate}
%After obtaining the initial prediction at level 4, we compute the residual error map, defined as the difference between the predicted mask and the downsampled ground truth mask at that scale. We up-sample the level 4 prediction after residual correction back to the resolution of level 3. At level 3, another XGBoost regression model is trained to predict the residual between the up-sampled mask and the ground truth mask, downsampled to level 3 resolution. We again concentrate on ROI voxels for this residual training. The predicted residual at level 3 is added to the upsampled mask to achieve a more accurate segmentation at that scale. This process is repeated for level 2 and then for level 1 (full $112\times112$ resolution): at each stage, the current prediction is interpolated to the next finer scale, and an XGBoost model corrects the discrepancies by learning the residual relative to the ground truth at that scale. By level 1, the model outputs a high-resolution probability map of the LV.

The overall architecture thus consists of a series of regression models across multiple scales, where the models have different targets: the initial coarse segmentation, boundary error correction at the coarse level, and upsampling error compensation at intermediate levels. MTGL’s design, which utilizes several small models in a hierarchy, results in a very lightweight segmentation pipeline compared to typical deep CNNs. 

\subsubsection{Classification decoder}\label{class-branch}

For the task of LVEF classification, we build on the feature representations learned by the unsupervised VoxelHop encoder (as described in Section~\ref{vox-encoder}). While the segmentation decoder provides the anatomical boundaries necessary for calculating ventricular volumes (EDV and ESV) according to the standard Simpson's method~\cite{mcgowan2003reliability}, the parallel classification decoder (XGBoost) predicts the clinical LVEF category directly from the deep feature representations. In particular, the multi-level VoxelHop encoder provides a rich set of features that capture spatiotemporal patterns in the echo volume, which can be repurposed for predicting the patient’s LVEF category. Our classification method uses these features in a two-step process: first pooling the features into a compact descriptor, then applying an XGBoost classifier to predict the LVEF category.

For each input data, we forward it through the four VoxelHop layers and collect the output feature maps at each level. Let $F_1, F_2, F_3, F_4$ denote the feature maps from level 1 (finest) through level 4 (coarsest), respectively, with $F_\ell \in \mathbb{R}^{H_\ell \times W_\ell \times T_\ell \times D_\ell}$ and we apply a different pooling strategy to each feature map:

\begin{enumerate}[1.]
\item For the finest level features $F_1$ and the coarsest level $F_4$, we apply Spatial Pyramid Pooling (SPP) \cite{spp}. SPP (Eq.~(\ref{eq1})) partitions the feature map into a set of spatial bins at multiple scales and computes pooled statistics within each bin, then concatenates these statistics. This allows us to retain multi-scale spatial information from those feature maps.

\begin{equation}
SPP=\bigl[\;\tfrac{1}{|\mathcal{B}_{u,v,w}|}\!\!\sum_{(i,j,t)\in\mathcal{B}_{u,v,w}} \!F_l(i,j,t,\!:\!)\;\bigr]_{u,v}
\quad\text{with a }2\times 2\times 1\ \text{SPP grid}.\label{eq1}
\end{equation}

\item For the intermediate level feature maps $F_2$ and $F_3$, we apply Global Average Pooling (GAP) \cite{GAP}, which computes the mean feature activation across the entire spatial map. GAP provides a compact summary of these feature sets by averaging out the spatial dimensions entirely (this is equivalent to an SPP with only a 1$\times$1 bin). We choose GAP (Eq.~(\ref{eq2})) for the middle levels to favor coarse global information, whereas SPP captures finer details and abstract global context at the extremes (very high-resolution and very low-resolution features), respectively, as given by

\begin{equation}
GAP=\Bigl[\tfrac{1}{H_1W_1T_1}\sum_{i,j,t}F_l(i,j,t,\!:\!),\ \max_{i,j,t}F_l(i,j,t,\!:\!)\Bigr]\label{eq2}.
\end{equation}

This pooling strategy yields four pooled feature vectors $h_1, h_2, h_3, h_4$, one per VoxelHop level. We then concatenate these vectors to form a single multi-level feature descriptor $H = [h_1 ; h_2 ; h_3 ; h_4]$ for each input volume. The dimensionality of $H$ is the sum of the lengths of the individual $h_i$ vectors (on the order of a few thousand, depending on $D_i$ and pooling configuration). Notably, by incorporating features from all four scales, $H$ contains information ranging from local motion/texture patterns (from $F_1$) to global shape and temporal dynamics (from $F_4$), which are all potentially relevant to assessing cardiac function.

\end{enumerate}

The concatenated feature vector $H$ is used as input to a supervised classifier to predict the LVEF category. We opted for an XGBoost classification model for this task. We train the XGBoost model using the corresponding EF class label for each data point. The model is optimized to minimize the multi-class classification error. 

By leveraging the unsupervised learned features from the MTGL encoder, our classifier benefits from representations that already capture the anatomical and motion characteristics of the LV. This approach is efficient: the encoder’s features are computed only once, and the classification model itself (XGBoost) is lightweight. Furthermore, because the segmentation mask information was embedded into the feature extraction process (recall that the VoxelHop encoder was trained on image and mask volumes), the features $H$ inherently incorporate knowledge of the LV region and boundaries. This effectively guides the classifier to focus on the relevant cardiac structures for predicting LVEF. In summary, the classification pipeline utilizes the interpretable, multi-scale features generated by the MTGL model to achieve accurate EF category prediction, thereby demonstrating cross-task transfer of unsupervised representational learning.

\subsection{Evaluation Metrics}\label{subsec2}

%To quantitatively assess the performance of our model, we employed distinct metrics for the segmentation and classification tasks. All metrics were computed for each case in the test set, and we report the aggregated mean values.

%\subsubsection{Segmentation Metrics}
To evaluate the accuracy of LV segmentation, we utilized the Dice Similarity Coefficient (DSC, Eq.~(\ref{DSC-metrice}) and the Intersection over Union (IoU, Eq.~(\ref{iou-metrices})). For the three-class LVEF classification task, we utilized the overall Accuracy (Acc, Eq.~(\ref{ACCEQ})) and Balanced Accuracy (BA, Macro Recall~\ref{BAEQ}) to evaluate the model's performance. All statistical analyses were conducted using Python 3.8 and SciPy 1.16.2.

\paragraph{Dice Similarity Coefficient (DSC)} The DSC (Eq.~(\ref{DSC-metrice})) quantifies the spatial overlap between the predicted segmentation mask ($A$) and the ground-truth mask ($B$). It is particularly sensitive to the agreement of the segmented regions. The coefficient is defined as:
\begin{equation}
    \text{DSC}(A, B) = \frac{2 \times |A \cap B|}{|A| + |B|}.\label{DSC-metrice}
\end{equation}
The DSC score ranges from 0, indicating no overlap, to 1, indicating a perfect match between the prediction and the ground truth.

\paragraph{Intersection over Union (IoU)} The IoU (Eq.~(\ref{iou-metrices})) quantifies the spatial overlap between the predicted segmentation mask ($A$) and the ground-truth mask ($B$) relative to their union. It penalizes both over- and under-segmentation by measuring the proportion of shared area compared with the total area covered by either mask. The index is defined as:
\begin{equation}
\text{IoU}(A, B) = \frac{|A \cap B|}{|A \cup B|} = \frac{|A \cap B|}{|A| + |B| - |A \cap B|}.\label{iou-metrices}
\end{equation}
The IoU score ranges from 0, indicating no overlap, to 1, indicating a perfect match between the prediction and the ground truth.

\subsubsection{Classification Metric}
For the three-class LVEF classification task, we used the overall Accuracy (Acc, Eq.~(\ref{ACCEQ})) to evaluate the model's performance.

\paragraph{Accuracy (Acc)} Accuracy represents the proportion of samples correctly classified among the total number of samples. It provides a direct measure of the classifier's overall correctness across the three LVEF categories. The formula is given by:
\begin{equation}\label{ACCEQ}
    \text{Accuracy} = \frac{\text{Number of Correctly Classified Samples}}{\text{Total Number of Samples}} = \frac{\sum_{i=1}^{C} \text{TP}_i}{N},
\end{equation}
where $\text{TP}_i$ is the number of true positives for class $i$, $C$ is the total number of classes (in our case, $C=3$), and $N$ is the total number of samples.

\paragraph{Balanced Accuracy (BAcc)}
Since the three LVEF classes are imbalanced, we additionally report balanced accuracy, also known as macro recall. Balanced accuracy is defined as the unweighted mean of the per class recall values, giving equal importance to each class regardless of its prevalence. For a $C$ class problem, let $\text{TP}_i$ and $\text{FN}_i$ denote the true positives and false negatives for class $i$. The recall for class $i$ is
\begin{equation}\label{RECEQ}
\text{Recall}_i = \frac{\text{TP}_i}{\text{TP}_i + \text{FN}_i}.
\end{equation}
Balanced accuracy is then
\begin{equation}\label{BAEQ}
\text{BA} = \frac{1}{C}\sum_{i=1}^{C} \text{Recall}_i
= \frac{1}{C}\sum_{i=1}^{C}\frac{\text{TP}_i}{\text{TP}_i + \text{FN}_i}.
\end{equation}
In terms of the confusion matrix $\mathbf{M}$ with entries $M_{i,j}$ (true class $i$ predicted as $j$), we have $\text{TP}_i = M_{i,i}$ and $\text{TP}_i + \text{FN}_i = \sum_{j=1}^{C} M_{i,j}$, so $\text{Recall}_i = M_{i,i}/\sum_j M_{i,j}$ and $\text{BA}$ is the mean of these $C$ recalls.

\section{Results}\label{sec2}

\subsection{MTGL outperforms deep learning benchmarks in a multi-task setting}
%==== Done======↓
The proposed MTGL framework demonstrates comprehensive superiority over contemporary 3D DL models in both LV segmentation and LVEF classification. As detailed in Table~\ref{tab1-n}, MTGL achieves the highest performance across all evaluated metrics on the EchoNet-Dynamic test set.

\begin{table}[h]
\caption{Comparative performance of MTGL against state-of-the-art 3D DL models on the EchoNet-Dynamic dataset. The proposed model demonstrates superior performance across all key metrics for both classification and segmentation tasks. Best results are highlighted in bold. Acc: Accuracy (Classification); BAcc: Balanced Accuracy (Classification); DSC: Dice Similarity Coefficient (Segmentation); IoU: Intersection over Union (Segmentation).}\label{tab1-n}%
\begin{tabular}{@{}lcccc@{}}
\toprule
Model & Acc & BAcc & DSC & IoU\\
\midrule

3D V-Net  \cite{vnet} & 0.7905 (-16.16\%) & 0.5177 (-35.9\%) & 0.8914 (-2.25 \%)  & 0.8045 (-4.03 \%)\\
3D UNETR  \cite{unetr} & 0.8476 (-10.11\%) & 0.5637 (-30.21\%) & 0.8604 (-5.65 \%) &  0.7623 (-9.07 \%) \\
3D U-Net \cite{unet} & 0.9048 (-4.04\%) & 0.7190 (-10.98\%) & 0.8846 (-3.00 \%)  &  0.7933 (-5.37 \%)\\
3D nnU-Net \cite{nnunet} & 0.9333 (-1.02\%) & 0.7460 (-7.64\%) & 0.9011 (-1.18 \%)  &  0.8204 (-2.14 \%) \\
\textbf{Ours}  & \textbf{0.9429}  & \textbf{0.8077} & \textbf{0.9119}  & \textbf{0.8383} \\
\botrule
\end{tabular}
\end{table}
%For the three-class LVEF classification task, the multi-task GUSL model attains an overall accuracy of 94.29\%. This represents a significant and clinically meaningful improvement over all deep learning baselines, including an absolute gain of +4.04\% over the strongest competitor 3D U-Net with 90.5\% overall accuracy. The performance gap is even more pronounced when compared to the transformer-based UNETR (84.76\%) and the 3D V-Net (79.05\%), which exhibit relative accuracy drops of 10.11\% and 16.16\%, respectively, compared to our model.

For the three-class LVEF classification task, the MTGL model achieves the highest accuracy of 0.9429, consistently outperforming state-of-the-art baselines. While standard models like 3D V-Net and UNETR exhibit substantial relative accuracy drops of 16.16\% and 10.11\%, respectively, our model also surpasses the highly competitive 3D nnU-Net by a margin of 1.02\%. To account for the imbalanced class distribution, we prioritized Balanced Accuracy as a key performance indicator. In this metric, the MTGL model demonstrates superior robustness, achieving a score of 0.8077. This marks a significant improvement over the comparative models, which suffer performance degradations ranging from 7.64\% (3D nnU-Net) to as much as 35.9\% (3D V-Net). These results indicate that unlike single-task baselines, the proposed multi-task approach does not bias toward the majority class and maintains high predictive power across all LVEF categories.

%In the concurrent segmentation task, Multi-task GUSL yields the highest Dice Similarity Coefficient (DSC) of 0.912, indicating superior spatial overlap with the ground-truth LV masks. This result surpasses the 3D V-Net (0.891), 3D U-Net (0.885), and UNETR (0.860) by absolute margins of +2.25\%, +3\%, and +5.65\%, respectively. Furthermore, Multi-task GUSL achieves the most precise boundary delineation, as evidenced by the lowest 95th percentile Hausdorff Distance (HD95) of 1.834. This metric, which quantifies contour accuracy, is reduced by 8.2\% relative to the 3D U-Net (1.998) and by a substantial 32.4\% relative to UNETR (2.714). These concurrent gains in both regional overlap (DSC) and boundary fidelity (HD95) confirm that the GUSL framework produces segmentations that are not only globally accurate but also locally precise at the critical endocardial border.

In the concurrent segmentation task, MTGL yields the highest DSC of 0.912 and IoU of 0.8383, indicating superior spatial overlap with the ground-truth masks. While standard architectures like 3D V-Net and UNETR lag behind with DSC deficits of 2.25\% and 5.6\% respectively, our model also outperforms the robust 3D nnU-Net baseline (DSC 0.9011) by 1.18\%. Qualitative inspection of the segmentation outputs confirms that MTGL produces masks that are both globally accurate and locally precise at the critical endocardial border, reflecting the effectiveness of its coarse-to-fine residual regression decoder. Unlike Self-Supervised Learning (SSL) methods that require massive datasets to learn generalized representations without overfitting, the VoxelHop encoder relies on PCA to derive filters, which stabilize much faster with fewer samples than the non-convex optimization landscape of deep neural networks.

%==== Done======↑
%==== Done======↓
\subsection{Superior segmentation fidelity with sharper boundary delineation}

%To further investigate the segmentation performance, we analyzed the distribution of DSC scores across the entire test set. As illustrated in Figure \ref{Seg-results}, the GUSL model not only achieves the highest median DSC but also exhibits significantly lower variability compared to the deep learning baselines. The tighter interquartile range and shorter whiskers of the GUSL boxplot indicate a more consistent and reliable segmentation performance, a crucial attribute for clinical applications where predictable accuracy is paramount. Statistical analysis confirms that the improvement in DSC for GUSL is highly significant compared to all baseline models (two-sided paired t-test, Bonferroni-corrected $p<10^{-4}$).

%To further investigate segmentation performance, the distribution of DSC and IoU scores across test subjects was analyzed. As shown in Figure \ref{Seg-results}, the GUSL model achieves the highest mean DSC and IoU with significantly lower variance compared to all baselines (paired t-test, Bonferroni-corrected $p<10^{-4}$). These results indicate that GUSL consistently outperforms alternative methods not only on average but also in reliability.

\begin{figure}[tbp]
  \centering

  \begin{subfigure}[t]{0.48\textwidth}
    \centering
    \includegraphics[width=\linewidth]{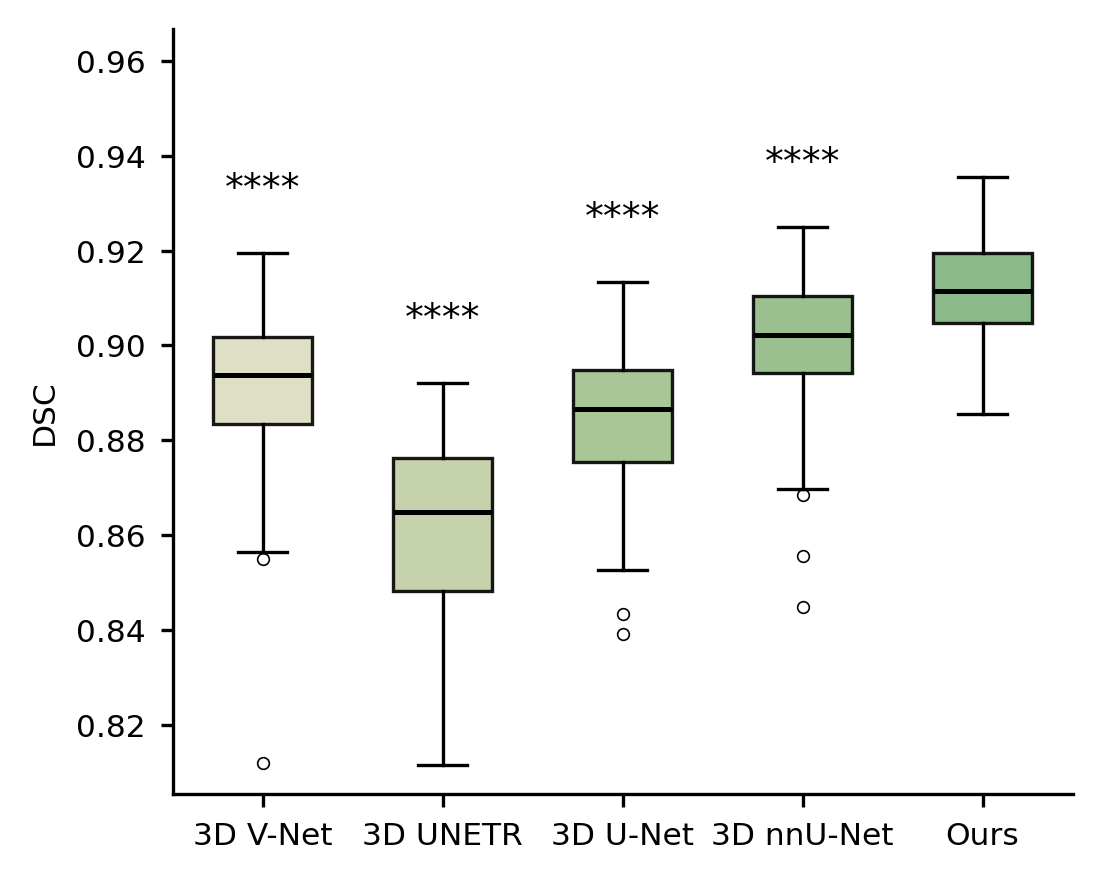}
    \caption{}
    \label{fig:seg-a}
  \end{subfigure}\hfill
  \begin{subfigure}[t]{0.48\textwidth}
    \centering
    \includegraphics[width=\linewidth]{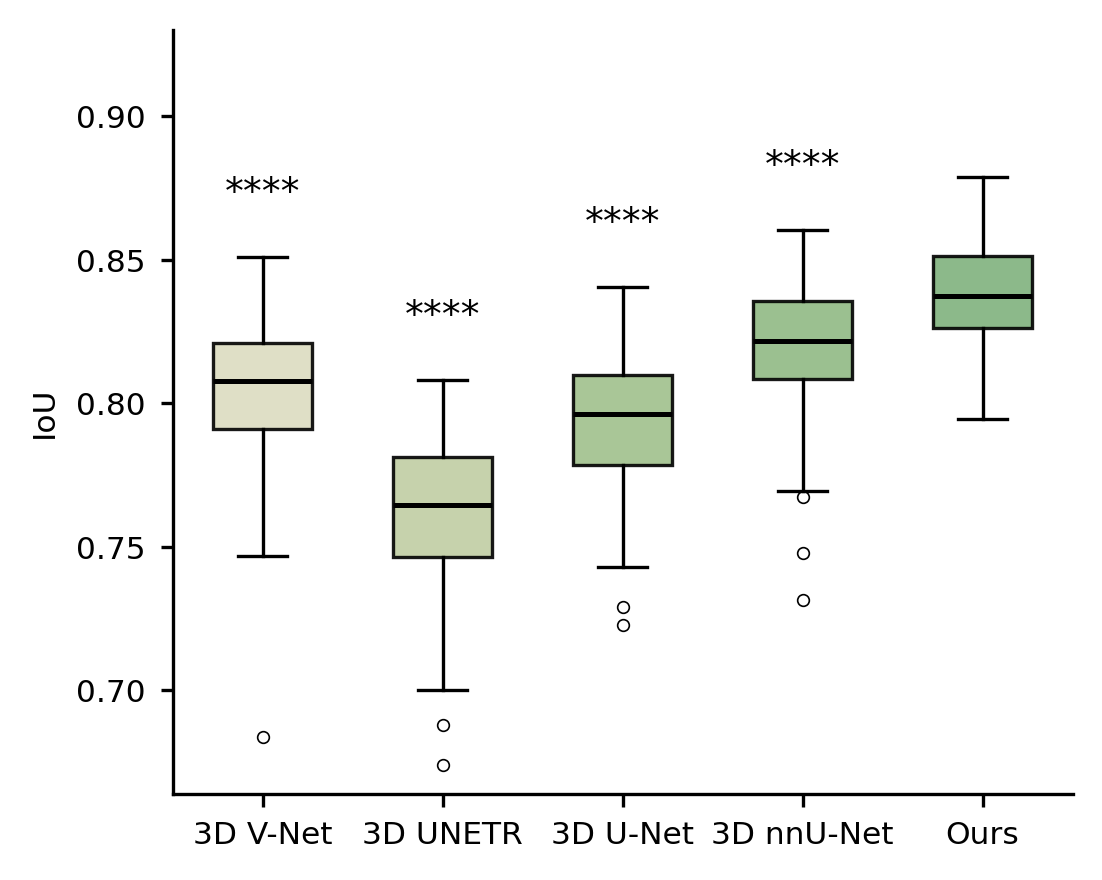}
    \caption{}
    \label{fig:seg-b}
  \end{subfigure}

  \caption{LV segmentation accuracy DSC (a) and IoU (b) across models. Boxplots show the median, interquartile range (IQR), and overall range for 3D V-Net, 3D UNETR, 3D U-Net, 3D nnU-Net, and the proposed MTGL model. Our method yields the highest median DSC and IoU with the lowest variability, indicating superior and more consistent performance. Asterisks denote statistical significance versus our model (two-sided paired t-test, Holm–Bonferroni corrected); **** denotes $p<10^{-4}$.}
  \label{Seg-results}
\end{figure}

To further investigate the segmentation performance, we analyzed the distribution of DSC scores across the entire test set. As shown in Figure~\ref{Seg-results}, the MTGL model achieves the highest mean DSC and IoU with significantly lower variance compared to all baselines (paired t-test, Bonferroni-corrected $p<10^{-4}$). These results indicate that MTGL consistently outperforms alternative methods not only on average but also in reliability.

%\begin{figure}[h]
%\centering
%\includegraphics[width=0.9\textwidth]{figure/dsc_boxplot.png}
%\caption{LV segmentation accuracy (Dice, DSC) across models. Boxplots show the median, interquartile range (IQR), and overall range for V-Net, UNETR, 3D U-Net, and the proposed GUSL model. Our method yields the highest median DSC with the lowest variability, indicating superior and more consistent performance. Asterisks denote statistical significance versus our model (two-sided paired t-test, Bonferroni-corrected); **** denotes $p<10^{-4}$.}\label{Seg-results}
%\end{figure}

% 內文

\begin{sidewaysfigure}
  \centering
 
  \includegraphics[width=0.9\textheight]{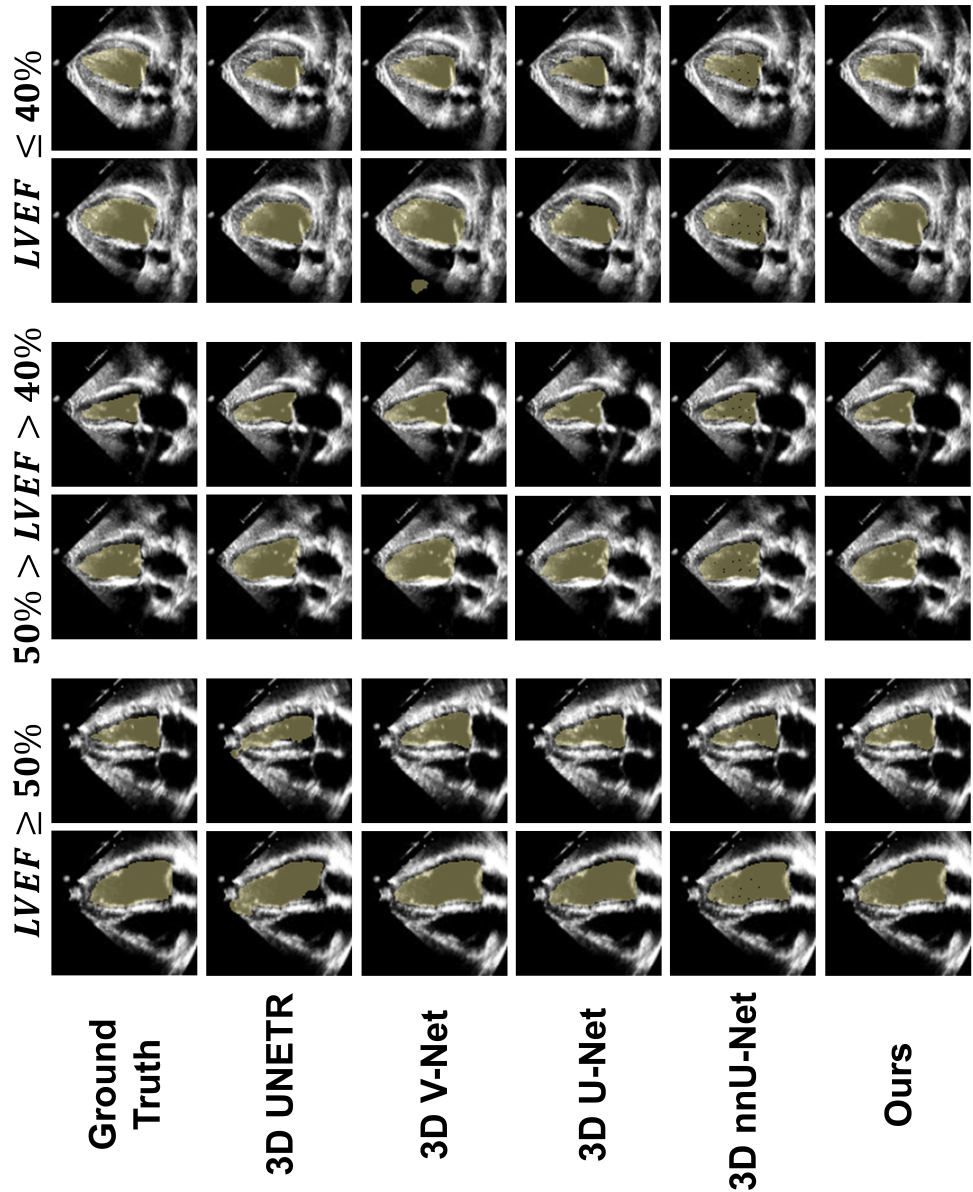}
  \caption{Representative segmentation results on example test cases with different LVEF ranges (columns) are shown. Ground-truth LV contours are compared with predictions from 3D UNETR, 3D V-Net, 3D U-Net, 3D nnU-Net and the proposed MTGL (Ours) model (rows). The MTGL segmentations closely match the true endocardial border across normal (LVEF $\geq 50\%$), borderline (40--50\%), and reduced (LVEF $\leq 40\%$) EF cases, whereas the deep CNN baselines exhibit larger errors (e.g., missed apical regions or spurious segments) in more challenging low-EF examples. These qualitative results consistently indicate high segmentation fidelity by MTGL, even in dilated, low-functioning ventricles, highlighting the model’s robustness across LVEF strata.}
  \label{seg-all}
\end{sidewaysfigure}

Qualitative evaluation of the segmentation outputs provides compelling visual evidence. Figure~\ref{seg-all} displays representative segmentation results on test cases with varying LVEF ranges. Across all categories, the MTGL segmentations consistently and accurately trace the true endocardial border. In contrast, the DL models frequently exhibit conspicuous errors, particularly in the more challenging low-LVEF cases, which often feature dilated and abnormally shaped ventricles. Common failure modes for the baseline models include missing the apical region of the ventricle or generating spurious, disconnected segments. The high fidelity of MTGL's segmentations, even in these difficult, low-functioning ventricles, highlights the model’s robustness across diverse clinical strata and emphasizes the effectiveness of its coarse-to-fine residual regression decoder.

%==== Done======↑

%==== Done======↓
\subsection{A paradigm of efficiency: Drastic reduction in model complexity}

%A defining advantage of the GUSL framework is its exceptional computational efficiency, a direct result of the Green Learning paradigm. As detailed in Table \ref{tab2}, our model achieves its state-of-the-art performance with only 1.13 million parameters. This represents a dramatic reduction in model complexity compared to the deep learning baselines. The 3D U-Net, UNETR, and 3D V-Net require 16.32M, 10.47M, and 45.61M parameters, respectively, making them 14.4, 9.5, and a staggering 40.4 times larger than our GUSL model.

%A defining advantage of the GUSL framework is its exceptional efficiency, which stems from its compact multi-hop architecture. The full multi-task model requires only about 1.13 million parameters (Table \ref{tab2}), whereas typical 3D convolutional networks use an order of magnitude more (tens of millions). For example, the baseline networks in Table \ref{tab2} contain approximately 9.5, 23.0, and 40.4 times more parameters than GUSL. This drastic shrinkage translates into tangible computational and environmental benefits at training time and at deployment. Under a standardized hardware configuration (Intel Xeon Gold 6342 CPU, 100 GB RAM, NVIDIA RTX A6000), GUSL attains competitive runtime while achieving the lowest energy use and carbon emissions among the compared methods (4.97 kWh and 2.09 kg CO\textsubscript{2}, respectively), compared with 3D U-net (5.50 kWh, 2.31 kg), 3D UNETR (5.04 kWh, 2.12 kg), and 3D V-net (7.74 kWh, 3.25 kg) as shown in Table \ref{train-carbon}.

A defining advantage of our MTGL framework is its exceptional efficiency, which stems from its compact multi-hop architecture. The full multi-task model requires only about 1.13 million parameters (Table~\ref{tab2}), whereas typical 3D convolutional networks use an order of magnitude more (tens of millions). For example, the baseline networks in Table \ref{tab2} contain approximately 9.5, 23.0, and 40.4 times more parameters than MTGL. This drastic shrinkage translates into tangible computational and environmental benefits at training time and at deployment. Under a standardized hardware configuration (Intel Xeon Gold 6342 CPU, 100 GB RAM, NVIDIA RTX A6000), MTGL attains competitive runtime while achieving the lowest energy use and carbon emissions among the compared methods (4.97 kWh and 2.09 kg CO\textsubscript{2}, respectively) as shown in Table \ref{train-carbon}.

\begin{table}[t]
\caption{Comparison of model complexity. MTGL contains substantially fewer parameters than the DL baselines, highlighting its computational efficiency and alignment with GL principles.}\label{tab2}%
\begin{tabular}{@{}lcc@{}}
\toprule
Model & Feature per dataset after pooling & \# of parameters (M)\\ 
\midrule
3D V-net  \cite{vnet} & 1184  & 45.61 \textbf{(40.36x)}  \\
3D UNETR  \cite{unetr} & 672  & 10.47 \textbf{(9.50x)} \\
3D U-net \cite{unet} & 2688  & 16.32 \textbf{(14.4x)}  \\
3D nnU-Net \cite{nnunet} & 1920  & 49.08 \textbf{(43.43x)} \\
\textbf{Ours}   & 3697  & 1.13  \\
\botrule
\end{tabular}
\end{table}

\begin{table}[t]
\caption{Comparison of carbon footprint and energy consumption in model training. Calculation based on using an Intel Xeon Gold 6342 2.8GHz 24 Core CPU with 100 GB Memory and an NVIDIA RTX A6000 48 GB GPU.}\label{train-carbon}%
\begin{tabular}{@{}lccc@{}}
\toprule
Model & Runtime & CO$_{2}$ (kg) & Energy (kWh)\\
\midrule

3D V-Net  \cite{vnet} & 8.75 Hours (CPU + GPU)  & 3.25  & 7.74 \\
3D UNETR  \cite{unetr} & 5.7 Hours (CPU + GPU)  & 2.12 & 5.04 \\
3D U-Net \cite{unet} & 6.22 Hours (CPU + GPU)  &  2.31  & 5.5 \\
3D nnU-Net \cite{nnunet} & 10.42 Hours (CPU + GPU)  &  3.87  & 9.22 \\
\textbf{Ours}  & \textbf{6.48 Hours (CPU + GPU)\footnotemark[1]}  & \textbf{2.09}  & \textbf{4.97} \\
\botrule
\end{tabular}
\footnotetext[1]{This configuration includes 1.52 hours of encoder training on the CPU and 4.96 hours of decoder training on both the CPU and GPU.}
\end{table}

%This drastic parameter reduction has profound practical implications. A smaller model translates to faster training times, lower energy consumption, and a significantly smaller carbon footprint. These characteristics are essential for real-world clinical deployment, particularly on resource-constrained edge devices such as portable ultrasound machines or mobile health platforms. While the GUSL model generates a relatively high-dimensional feature vector, the subsequent classification is performed by a lightweight and efficient XGBoost tree ensemble. This is in stark contrast to DL models, which typically employ massive, computationally expensive fully-connected layers or attention mechanisms for their final prediction heads. The efficiency of GUSL thus demonstrates a more effective use of parameters, leveraging feature richness rather than sheer weight count to achieve superior accuracy, which is a core tenet of our GL philosophy.

%==== Done======↓
\subsection{Opening the black box: Interpretable features for clinical trust}

Beyond performance and efficiency, our MTGL framework offers inherent interpretability, a critical feature for building clinical trust and a key differentiator from opaque ``black box" models. Because the VoxelHop encoder is constructed from a series of linear PCA transformations, its learned features can be directly visualized and understood.

An ablation study was carried out to evaluate the contribution of the hierarchical VoxelHop feature layers to LVEF classification, with the results presented in Table \ref{tab3}. Combining features from all hops yields the best accuracy ($0.9429$ at $3697$ pooled features), indicating that multi-scale information from all levels is complementary. Excluding only the \textit{Hop~1} results in a very small accuracy drop (to 93.33\%), while using only the two deepest hops (\textit{Hop~3} and \textit{Hop~4}, totaling 3,387 features) still achieve a strong 91.43\% accuracy—nearly matching the full feature set, highlighting that high-level spatio-temporal abstractions are the most discriminative for LVEF grouping. In contrast, using only the shallowest hops (\textit{Hop~1} or \textit{Hop~1}+\textit{Hop~2}) caused a pronounced accuracy decline (down to~83–84\%), implying that local texture and edge cues are insufficient without deeper, context-rich features. Notably, \textit{Hop~4} alone achieves  91.43\% accuracy with $2,960$ features, emphasizing that the deepest hop captures substantial class-relevant variation; however, the full-hop combination closes the remaining gap and delivers the best result.

\begin{table}[h]
\caption{Ablation study on the contribution of different VoxelHop feature layers to Left Ventricular Ejection Fraction (LVEF) classification. The results quantify the importance of the hierarchical features, showing that while deeper features (\textit{Hop~3} and \textit{Hop~4}) are highly discriminative, combining features from all levels achieves the optimal outcome.}\label{tab3}

\begin{tabular}{@{}lcccccc@{}}
\toprule
Ranking & \textit{Hop~1} & \textit{Hop~2} & \textit{Hop~3} & \textit{Hop~4} & Feature & Acc \\
\midrule
1 & + & + & + & + & 3697 & \textbf{0.9429} \\
2 & - & + & + & + & 3589 & 0.9333 \\
3 & - & - & + & + & 3387 & 0.9143 \\
4 & - & - & - & + & 2960 & 0.9143 \\
5 & + & + & + & - & 737 & 0.8476 \\
6 & + & - & - & - & 108 & 0.8381 \\
7 & + & + & - & - & 310 & 0.8190 \\

\botrule
\end{tabular}
\end{table}

%==== Done======↑

%==== Done======↓

\begin{figure}[h]
\centering
\includegraphics[width=0.9\textwidth]{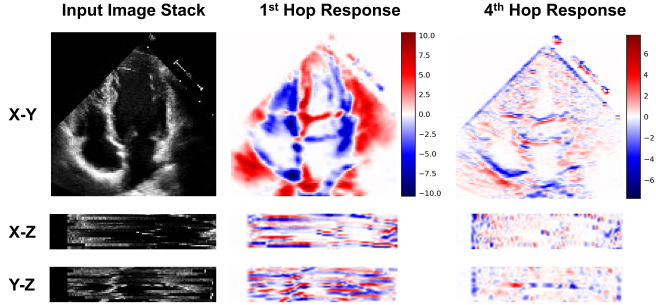}
\caption{Interpretable multi-hop feature responses: Response of the most influential AC filter in the 1st and 4th hops of the VoxelHop encoder. The color bar visualizes the AC filter response computed on z-score standardized voxel intensities (arbitrary units, a.u.). Red indicates positive filter responses (stronger activation) and blue indicates negative responses (suppression), with the magnitude given by the color bar.}\label{fig-AC-filter}
\end{figure}

Figure~\ref{fig-AC-filter} illustrates the hierarchical feature extraction process. Let $\mathcal{I}(\mathbf{x})$ denote the input volume, where $\mathbf{x}=(x,y,z)$ represents spatiotemporal coordinates. The response map $R^{(k)}_l$ at layer $l$ is formally defined as the projection of the local neighborhood onto the principal subspace (AC kernel $\mathbf{K}^{(k)}_l$):
\begin{equation}
    R^{(k)}_l(\mathbf{x}) = (\mathcal{I} * \mathbf{K}^{(k)}_l)(\mathbf{x}) = \sum_{\mathbf{u} \in \Omega} \mathbf{K}^{(k)}_l(\mathbf{u}) \, \mathcal{I}(\mathbf{x}-\mathbf{u}).
\end{equation}
By visualizing the most energetic kernels from the shallowest (\textit{Hop~1}) and deepest (\textit{Hop~4}) layers, we observe a transition from local edge detection to global dynamic analysis.
%Figure~\ref{fig-AC-filter} illustrates the responses obtained by convolving a representative echo volume \(I(x,y,z)\) with the most influential AC kernels from the shallowest (\textit{Hop~1}) and deepest (\textit{Hop~4}) encoder layers: \(R(\mathbf{p})=(I*K)(\mathbf{p})=\sum_{\mathbf{u}}K(\mathbf{u})\,I(\mathbf{p}-\mathbf{u})\), where the color map is zero–centered (blue/negative, red/positive). We visualize three orthogonal slices, including spatial X–Y, and spatiotemporal X–Z and Y–Z (with \(z\) indexing the interleaved End-Diastolic Volume (EDV)/End-Systolic Volume (ESV) frames), to disentangle what each hop is sensitive to. This visualization illustrates a logical, hierarchical feature extraction process that progresses from local texture to global dynamics. 

\textbf{X–Y (spatial):} 

The learned kernel at \textit{Hop~1} exhibits high correlation with a first-order gradient operator ($\mathbf{K}^{(1)} \approx \nabla_{xy}$). Consequently, its response forms biphasic ridges that straddle the endocardial border: narrow negative–positive lobes that straddle the endocardial border and the mitral–annular line, with additional paired responses at papillary muscle heads and chordal insertions. In fact, the learned AC kernel shows near–perfect correlation with a first–order template, so its response is well-approximated by a directional derivative of a locally smoothed image. Therefore, the sign of the lobe pair encodes the orientation of the blood–myocardium transition (bright cavity to darker myocardium), producing red (positive) on the myocardial side and blue (negative) on the luminal side when it points from cavity to wall. High–magnitude ridges along the septum and lateral wall delineate the LV cavity with subvalvular continuity toward the apex. Thinner, punctate dipoles appear over papillary muscles and chordae where small, highly echogenic structures disrupt the intracavitary intensity plateau. At the valve plane, leaflet edges and annulus generate thin, arcuate dipoles consistent with specular reflections from fibrous tissue; these crisp spatial cues explain why shallow hops are essential for segmentation. 
In contrast, \textit{Hop~4} shows low–amplitude, spatially diffuse activity in X–Y. Intuitively, its learned AC kernel behaves like a derivative through the stack rather than within the plane, so it suppresses purely in–plane contrast and only leaves mild activity near the base where through–plane motion and leaflet kinematics leak into the slice. In summary, \textit{Hop~4} is less sensitive to the contour location in a single frame and more sensitive to how the image changes between EDV and ESV frames (spatiotemporal dynamics).

%==== Done======↑

%==== Done======↓

\textbf{X–Z / Y–Z (spatiotemporal):}\\
Viewed in longitudinal (X–Z, Y–Z) slices, the two hops separate cleanly in role. \textit{Hop~1} forms thin stripes that follow the moving endocardial border and leaflet tips across \(z\) (EDV–ESV alternation). These tracks mark the locations of tissue interfaces at each frame and remain sharp over time, providing anatomically faithful cues at the septum, lateral wall, and mitral annulus.

In contrast, \textit{Hop~4} condenses the dynamics into broad, coherent red/blue bands along \(z\). Because it approximates a temporal difference operator ($R \propto \partial \mathcal{I} / \partial z$), positive lobes appear where intensity systematically increases from EDV to ESV (wall thickening, cavity darkening), and negative lobes where it decreases (relaxation). The strongest bands occur at the valve plane and basal septum/lateral wall, reflecting annular descent and base–to–apex shortening (MAPSE motion) \cite{stoylen2021physiological, stoylen2023regional}, with sign flips at end–systole/diastole consistent with valve opening/closure. In other words, \textit{Hop~4} summarizes how much the ventricle changes between EDV and ESV, which precisely is the global contractile signal that drives LVEF discrimination, while \textit{Hop~1} preserves the high–frequency edges that anchor the geometry.

In a nutshell, \textit{Hop~1} explains “where the borders and small structures are” (valve leaflets, papillary muscles, chordae), enabling accurate contours; \textit{Hop~4} explains “how the heart moves overall” (annular descent, longitudinal shortening, cavity size swing), which is the class–discriminative signature for LVEF, and this finding is also aligned with the ablation study in Table \ref{tab3}.

%==== Done======↑

%==== Done======↓

\subsection{Accurate LVEF classification driven by clinically relevant feature learning}

Recognizing that the borderline LVEF category is both the most underrepresented and often the most clinically ambiguous, a targeted oversampling of this class was performed. As shown in Table~\ref{aug}, by augmenting the training data to balance the representation of Class 1, the model's accuracy surged to 94.3\%. This result indicates that the performance bottleneck was not the discriminative power of the VoxelHop features but rather the insufficient representation of this critical minority class. Further balancing of Class 2 conferred no additional benefit, confirming that restoring the representation of the borderline group was the pivotal step. This analysis demonstrates the model's robustness and its capacity to capitalize on its powerful feature set once the training data reflects the full spectrum of clinical presentations.

\begin{table}[h]
\caption{Effect of targeted oversampling on three-class Left Ventricular Ejection Fraction (LVEF)  classification. Values under Class 1/2/3 are sample counts per class after augmentation, Class 1 denotes as LVEF $>$ 50\%, Class 2 denotes as LVEF between 40–50\%, and Class 3 denotes as LVEF between $<$ 40\%.}\label{aug}%
\begin{tabular}{@{}lcccc@{}}
\toprule
Augmentation & Class 1 & Class 2 & Class 3 & Acc\\
\midrule
- & 482 & 60 & 79 & 0.8952 \\
Class 2 & 482 & 480 & 79 & 0.9429 \\
Class 3 & 482 & 60 & 480 & 0.8952 \\
Class 2 \& 2 & 482 & 480 & 480 & 0.9429 \\

\botrule
\end{tabular}

\end{table}

%The collective evidence indicates that GUSL’s feed-forward representation plus residual refinement improves both region overlap and boundary precision with substantially fewer parameters. The hop ablation confirms that deeper multi-scale descriptors carry most of the discriminative load for EF categorization, while shallow features provide complementary gains when combined. Class rebalancing analyses emphasize the importance of adequately sampling the borderline EF category to fully realize the benefits of the pooled VoxelHop features. Overall, the results support the claim that a white-box, low-parameter pipeline can match and surpass heavier 3D networks for joint cardiac segmentation and three-class EF classification.

\subsection{Parameter Selection - Energy Preserving}

Following the design philosophy of PixelHop \cite{pixelhop} and VoxelHop \cite{voxelhop}, which advocates for preserving as much signal energy as possible to retain task-relevant variation, the number of AC filters at each hop was selected according to the cumulative-energy criterion. Specifically, we sort the AC eigenvalues within a hop in descending order (after removing the DC term), compute the cumulative energy ratio, and choose the smallest value $K$ such that the ratio exceeds $99\%$. To avoid a brittle cutoff, a small safety margin was kept beyond that value of $K$. Under an orthogonal transform (Saab/PCA), discarding ACs corresponds to discarding their eigen–energies, thus a $99\%$ retention guarantees that the hop–wise relative $L_2$ reconstruction error is at most $1\%$, while the added safety margin further stabilizes downstream performance.

\begin{figure}[h]
\centering
\includegraphics[width=0.9\textwidth]{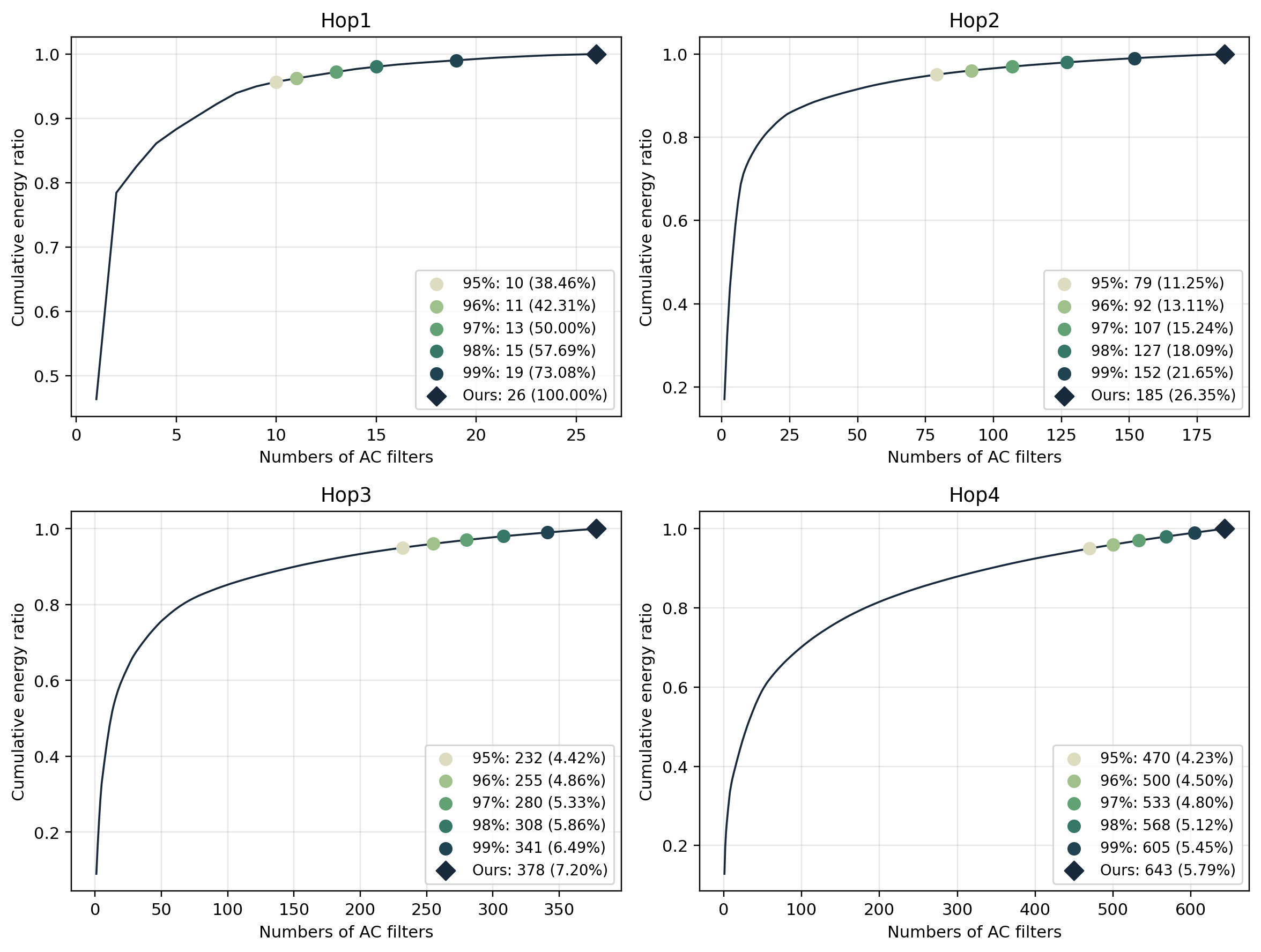}
\caption{The energy plot as a function of the number of AC filters. All numbers in this plot are computed from the saved spectra of the trained encoder, by concatenating per–node AC eigenvalues within each hop after DC removal and then applying the cumulative–energy rule. }\label{fig-energy}
\end{figure}

%(number of TH1–expanded local vectors \(\times\) \(3\!\times\!3\!\times\!3-1=26\))

Figure~\ref{fig-energy} visualizes the cumulative energy curves and marks the $95$–$99\%$ operating points together with our final kept AC counts (“Ours”). These choices place us safely to the right of the $99\%$ knee on every hop, which is consistent with \cite{pixelhop,voxelhop} and with the strong empirical accuracy we observe.

Although every hop we chose retains at least 99\% of energy, the final models are highly compact in terms of the number of AC filters. Moreover, the kept fraction of \textit{Hop~3} and \textit{Hop~4} is 7.20\% and 5.79\%, demonstrating that most high–order ACs contribute negligible energy and can be pruned without measurable loss. Aggregated over all hops, the AC budget is reduced from $17{,}082$ to $1{,}232$ filters (92.79\% reduction). 
%If we instead report against “active–nodes” (only those local vectors that still contain any AC after \texttt{TH2}), the global kept fraction is $1{,}232/8{,}086=\mathbf{15.24\%}$, leading to the same qualitative conclusion. Both normalizations are white–box, data–dependent, and directly interpretable from the spectra.
The cumulative–energy plots (Figure~\ref{fig-energy}) and table~\ref{hoplevel}expose where each hop’s information concentrates, and our selected points sit at the knee of the curve with explicit numerical guarantees, demonstrating the transparency of the VoxelHop encoder.  

\begin{table}[h]
\caption{Cumulative energy-based selection of AC filters per hop. Columns 95–99\% list the minimal number of AC filters \(K\) required to reach the indicated cumulative energy ratio within each hop (after DC removal). “Ours’’ is the number actually kept (chosen as \(K_{99}\) with safety margin). Percentages in parentheses are fractions of the hop’s potential ACs, where Total is the all–nodes baseline.}\label{hoplevel}
\begin{tabular}{@{}lccccccc@{}}
\toprule
\textit{Hop Level} & 95\% & 96\% & 97\% & 98\% & 99\% & Ours & Total\\
\midrule
\textit{Hop~1} & 10 (38.46 \%) & 11 (42.31\%) & 13 (50.00\%) & 15 (57.69\%) & 19 (73.08\%) & 26 (100.00\%) & 26 \\
\textit{Hop~2} & 79 (11.25 \%) & 92 (13.11\%) & 107 (15.24\%) & 127 (18.09\%) & 152 (21.65\%) & 185 (26.35\%) & 702\\
\textit{Hop~3} & 232 (4.42 \%) & 255 (4.86\%) & 280 (5.33\%) & 308 (5.86\%) & 341 (6.49\%) & 378 (7.20\%) & 5252\\
\textit{Hop~4} & 470 (4.23 \%) & 500 (4.50\%) & 533 (4.80\%) & 568 (5.12\%) & 605 (5.45\%) & 643 (5.79\%) & 11102\\

\botrule
\end{tabular}

\end{table}

\section{Discussion}\label{DIS}
This study introduces a MTGL framework that not only establishes a new state-of-the-art in automated echocardiographic analysis but also fundamentally challenges the prevailing paradigm in medical AI, which has long equated performance with model complexity. This confluence of superior accuracy and radical efficiency subverts the ``bigger is better" ethos that has dominated the DL landscape. A defining advantage of the MTGL framework, beyond its performance and efficiency, is its inherent transparency. This directly addresses the 'black box' problem, which remains one of the most significant barriers to the widespread clinical adoption, regulatory approval, and ultimate trustworthiness of AI systems in medicine \cite{quinn2022three}. In the MTGL model, interpretability is not a post-hoc approximation applied to an opaque system; it is a fundamental and inextricable property of its design.

The visualization of the VoxelHop encoder's feature responses, shown in Figure~\ref{fig-AC-filter}, provides a clear window into the model's decision-making process. This represents a profound departure from conventional DL, where explainability methods, such as saliency maps, often produce noisy, difficult-to-interpret heatmaps that highlight what the model is looking at, but fail to explain what it is seeing or why it is important. In our MTGL, features constitute the explanation, and such transparency is indispensable for clinical deployment. An interpretable model enables clinicians to understand the basis for a given prediction, which is crucial for building trust, identifying potential failure modes, and adjudicating outlier cases or unexpected results.  If a conventional DL model produces an erroneous segmentation, the cause is buried within the complex interplay of millions of learned weights, making targeted correction nearly impossible beyond retraining with more data. In contrast, if our MTGL model errs, the cause can be systematically investigated.

MTGL's lightweight, reliable characteristics enable it to run on edge devices, which could democratize expert-level cardiac assessment. This would empower a wider range of healthcare providers to perform rapid and accurate initial screenings for cardiac dysfunction, facilitating earlier diagnosis, more effective triage, and timely referral to specialist care. In high-volume echocardiography labs, such a tool could automate laborious and repetitive tasks, freeing up highly trained cardiologists and sonographers to focus on more complex diagnostic challenges and direct patient interaction, thereby enhancing overall workflow efficiency.

In the GL paradigm, compared with recent generative approaches. The VoxelHop encoder relies on the Saab transform, a linear subspace approximation method. While effective for capturing dominant spectral components of cardiac motion, this linearity may restrict the model's ability to represent highly complex, non-linear temporal dynamics or fine-grained motion anomalies that are better modeled by probabilistic diffusion models. Diffusion-integrated methods have demonstrated superior capabilities in modeling the non-linear manifold of echocardiographic video, particularly for tasks such as video synthesis or de-hazing. However, our results suggest that for the specific discriminative tasks of segmentation and LVEF classification, the linear spatiotemporal features extracted by MTGL are sufficiently robust, offering a pragmatic balance between modeling power and computational transparency.

Although our primary benchmarking focuses on representative 3D convolutional and transformer-based architectures trained under matched supervision on EchoNet Dynamic, we acknowledge that recent literature has reported strong performance using promptable foundation models and generative video segmentation approaches. For example, SAM variants~\cite{chao2025foundation} have reported the best overall DSC of 0.911 (with 0.929 at EDV and 0.894 at ESV). These results are important, but they are not directly comparable to a fully automatic, single-dataset, matched supervision evaluation because such methods may depend on large-scale external pretraining, interactive prompts, or substantially different learning objectives and evaluation protocols. In contrast, our approach estimates both EDV and ESV segmentations within a single multi-task pipeline, supporting phase-specific measurements without the need to train separate phase-specific models. We therefore present our results as a controlled comparison within a consistent experimental setting, while positioning foundation model and generative approaches as valuable complementary baselines for future work.

Despite these promising implications, it is essential to acknowledge the limitations of the present study and outline a clear path toward clinical translation. First, our validation was conducted on a single, albeit large and publicly available dataset: EchoNet-Dynamic \cite{ouyang2020video}. The robustness and generalizability of the model must be rigorously tested on external, multi-center datasets that capture the full spectrum of variability arising from different ultrasound vendors, patient demographics, and real-world acquisition protocols. Second, while our work establishes technical validity, the true clinical utility and impact of the MTGL framework can only be determined through prospective clinical trials. Third, the current analysis does not explicitly stratify performance based on image quality, a critical factor in clinical practice. Fourth, while the model is multi-task, its scope is currently limited to LV segmentation and three-class LVEF classification. Finally, while our analysis demonstrates reduced model complexity and favorable training-energy characteristics, we emphasize that real-world point-of-care deployment requires direct measurement on target hardware. We therefore temper deployment claims and explicitly identify edge benchmarking under realistic acquisition settings as a necessary next step. A truly comprehensive clinical tool would need to encompass a broader range of assessments, including the analysis of other cardiac chambers, valvular function, diastolic parameters, and myocardial strain areas.

%==== Done======↑
\section{Conclusion}\label{conclusion}

In this work, we have successfully developed and validated a multi-task, backpropagation-free multi-task Green Learning framework for the automated analysis of echocardiography. Our results demonstrate that this framework comprehensively surpasses state-of-the-art DL models, including 3D U-Net~\cite{unet}, 3D V-Net~\cite{vnet}, 3D UNETR~\cite{unetr}, and 3D nn-Unet~\cite{nnunet} in both LV segmentation and LVEF classification. The MTGL model not only leads in classification accuracy ($94.3\%$) and segmentation overlap (DSC $0.912$), but also achieves a dramatic model compression with only $1.13$M parameters—a reduction of over 10- to 40-fold compared to DL counterparts. This highlights a dual advantage in both performance and computational efficiency. Beyond performance and efficiency, the interpretability of the MTGL framework is a core advantage that distinguishes it from conventional black box models. By visualizing the features learned by the VoxelHop encoder, we can clearly trace the model's hierarchical abstraction from low-level edge details to high-level global dynamics.

\section*{Competing interests}
The authors declare no competing interests.

\section*{Funding}
%This research received no specific grant from any funding agency in the public, commercial, or not-for-profit sectors.
We acknowledge the partial funding support from the Graduate Institute of Biomedical Electronics and Bioinformatics, National Taiwan University.
\section*{Data availability}
%All data used in this study can be downloaded from the EchoNet-Dynamic \cite{ouyang2020video} dataset repository at \url{https://echonet.github.io/dynamic/index.html}.
All data used in this study can be downloaded from the EchoNet-Dynamic dataset~\cite{ouyang2020video}.

%\newpage





\bibliography{sn-bibliography}% common bib file

\begin{comment}

\newpage
\begin{table}[h]
\caption{Three-class LVEF classification performance on the EchoNet-Dynamic test set.}\label{class-balance}%
\begin{tabular}{@{}lcc@{}}
\toprule
Model & Accuracy\footnotemark[1] & Balanced Accuracy\footnotemark[1] \\
\midrule

3D V-Net  \cite{vnet} & 0.7905 (-16.16\%)    & 0.5177 (-35.9\%)\\
3D UNETR  \cite{unetr} & 0.8476 (-10.11\%)   &  0.5637 (-30.21\%)\\
3D U-Net \cite{unet} & 0.9048 (-4.04\%)    & 0.7190 (-10.98\%)\\
3D nnU-Net \cite{nnunet} & 0.9333  (-1.02\%) & 0.7460 (-7.64\%) \\
\textbf{Ours}    & \textbf{0.9429}  & \textbf{0.8077} \\
\botrule
\end{tabular}
\footnotetext[1]{Accuracy reflects overall performance under the natural class distribution, whereas balanced accuracy (macro recall) mitigates class-imbalance effects by equally weighting each class.
}
\end{table}
\end{comment}

\end{document}